\documentstyle[floats,aps]{revtex}
\advance\topmargin 30pt
\def\ut#1{#1\llap{\lower2ex\hbox{$\widetilde{\hphantom{#1}}$}}}
\begin{document}
\title{Nonstandard optics from quantum spacetime}
\author{Rodolfo Gambini$^{1}$\footnote{Associate member of ICTP.}, 
Jorge Pullin$^2$}
\address{1. Instituto de F\'{\i}sica, Facultad de Ciencias, 
Igu\'a 4225, esq. Mataojo, Montevideo, Uruguay}
\address{
2. Center for Gravitational Physics and Geometry, Department of
Physics,\\
The Pennsylvania State University, 
104 Davey Lab, University Park, PA 16802}
\date{September 4th 1998}
\maketitle
\begin{abstract}
We study light propagation in the picture of semi-classical space-time
that emerges in canonical quantum gravity in the loop
representation. In such picture, where space-time exhibits a
polymer-like structure at microscales, it is natural to expect
departures from the perfect non-dispersiveness of ordinary vacuum. We
evaluate these departures, computing the modifications to Maxwell's
equations due to quantum gravity, and showing that under certain
circumstances, non-vanishing corrections appear that depend on the
helicity of propagating waves. These effects could lead to observable
cosmological predictions of the discrete nature of quantum
spacetime. In particular, recent observations of non-dispersiveness in
the spectra of gamma-ray bursts at various energies could be used to
constrain the type of semi-classical state that describes the universe.
\end{abstract}

\vspace{-8.5cm} 
\begin{flushright}
\baselineskip=15pt
CGPG-98/9-1  \\
gr-qc/9809038\\
\end{flushright}
\vspace{8cm}

\bigskip

The recent discovery of the cosmological nature of gamma-ray bursts
opens new possibilities to use them as a laboratory to test
fundamental physics. This has been emphasized by Amelino-Camelia et
al. \cite{Amelino}. What these authors point out is that the light
coming from gamma-ray bursts travels very large distances before being
detected on Earth, and is therefore quite sensitive to departures from
orthodox theories. In particular, the bursts present detailed time
structures, with features smaller than $1 ms$, that are received
simultaneously through a broad band of frequencies, ranging from
$20keV$ to $300keV$, as reported by the BATSE detector of the Compton
Gamma Ray observatory \cite{milli}. This implies stringent limits on
any dispersive effects that light might suffer in travel towards the
Earth.

Various models of string quantum gravity imply dispersive frequency
wavelength relations for light propagation, and in reference 
\cite{Amelino} it was shown that the simultaneity of time structures in
the patterns of light received gamma ray bursts are possible
candidates to set limits on these models. In this note we would like
to probe similar issues for loop quantum gravity. An attractive
feature of this approach is that it might imply a unique signature of the
discrete nature of space time tantamount to an ``intrinsic
birefringence'' of quantum space-time. This effect would imply a
distinctive ``doubling'' of patterns observed in the time series
analysis of the bursts, making it attractive from the observational
point of view. We will see however, that the nature of the effects
predicted by loop quantum gravity depend on the type of semi-classical
state that one considers. In a sense, one can turn the argument around
and suggest that rather than viewing these effects as a prediction of
the theory, they can be used to constrain the type of semi-classical
states one considers to represent realistic cosmologies.

Loop quantum gravity \cite{rovelli} is usually formulated in the
canonical framework. The states of the theory are given by functions
of spin networks, which are a convenient label for a basis of
independent states in the loop representation. This kinematic
framework is widely accepted throughout various formulations of the
theory, and has led to several physical predictions associated with
the ``polymer-like'' structure of quantum space-time \cite{AsLe}. For
instance, a quite clear picture of the origin of the black hole
entropy emerges \cite{AsBaCoKr}. The dynamics of the theory is
embodied in the Hamiltonian constraint, and consistent proposals are
currently being debated \cite{GaLeMaPu}. To show the existence of the
birefringent effect we will not need too many details of the dynamics
of the theory. We prefer to leave the discussion a bit loose,
reflecting the state of the art in the subject, since there is no
agreement on a precise dynamics. Also, the spirit of our calculation
is to attempt to make contact with observational predictions,
something that is importantly lacking in the canonical approach, in
part as a consequence of the absence of a detailed prescription for
constructing the semi-classical limit of the theory. One should
therefore view the current work as a further elaboration towards
probing the nature of the semi-classical limit. Initial explorations
on this subject can be found in reference \cite{weave}.

The term in the Hamiltonian constraint coupling Maxwell
fields to gravity is the usual ``$E^2+B^2$'' term, but in a curved
background,
\begin{equation}
H_{\rm Maxwell} = {1 \over 2}\int d^3x \ut{g}_{ab} 
(\tilde{e}^a \tilde{e}^b+
\tilde{b}^a \tilde{b}^b).
\end{equation}
where we have denoted with tildes the fact that the fields are vector
densities in the canonical framework. This requires the division by
the determinant of the metric, which we denoted by an undertilde in
the metric. Thiemann \cite{Th} has a concrete proposal for realizing
in the loop representation the operator corresponding to the metric
divided by the determinant.

Since we are interested in low-energy, semi-classical effects, we will
consider an approximation where the Maxwell fields are in a state that
is close to a coherent state. That is, we will assume that the Maxwell fields
operate as classical fields at the level of equations of motion,
however, we will be careful when realizing the Hamiltonian to regulate
operator products.  This departs
from the regularization proposed by Thiemann. In his approach, the
states considered are such that the electric field operator is also
discrete and finite and therefore products at the same point are
acceptable. One can consider this as a feature of the full
diffeomorphism-invariant context, that will disappear at an effective
level when one considers semi-classical states. There one would expect
to recover the usual Maxwell theory with its divergences. The coherent
state chosen will be one that approximates a classical travelling wave
of wavelength $\lambda$, which we assume to be much larger than the
Planck length. 

For the gravitational degrees of freedom we
will assume we are in a ``weave'' state \cite{weave} $|\Delta>$, such
that,
\begin{equation}
\label{weave}
<\Delta|\hat{\ut{g}}_{ab}|\Delta> = \delta_{ab} + O({{\ell_P\over
\Delta}}),
\end{equation}
where $\ell_P$ is Planck's length.  Weave states \cite{weave},
characterized by a length $\Delta$, are constructed by considering
collections of Planck-scale loops. They are meant to be semi-classical
states such that that if one probes these states at lengths much
smaller than $\Delta$ one will see features of quantum space-time,
whereas if one probes at scales of the order of, or bigger than
$\Delta$ one would see a classical geometry. The weave we will 
consider approximates a flat geometry for lengths larger than
$\Delta$. It is worthwhile noticing that weave states were introduced
some time ago in the context of the loop representation, before a
variety of new techniques (cylindrical functions, spin networks) were
introduced to deal with the quantum states in this representation. At
the moment there is not a complete picture of how to construct weave
states in the loop representation in terms of spin network
states. When they were originally introduced, weave states were meant
to yield semi-classical behaviors in certain operators capturing
metric information of spacetime. It was evident that there were many
inequivalent states that could fit these requirements. If the reader
wishes, this paper introduces further requirements that we need to
demand from such semi-classical states. We will return to this
issue after we introduce the effects we wish to discuss.

Let us now consider the action of the Hamiltonian we proposed above on
a weave state. We need a few more details of the regularization of $
\ut{g}_{ab}$ that was proposed by Thiemann \cite{Th}. It consists in
writing $ \hat{\ut{g}}_{ab}$ as the product of two operators
$\hat{w}_a(x)$, each corresponding to a commutator of the Ashtekar
connection with the square root of the volume operator. The only
feature we will need of these operators is that acting on spin
network states they are finite and only give contributions at
intersections.  We now point split the operator as suggested in
\cite{Th}, (to shorten equations we only consider the electric part of
the Hamiltonian, the magnetic portion is treated in the same way)
\begin{equation}
\hat{H}^E_{\rm Maxwell}   = {1 \over 2} 
\int d^3x \int d^3y \hat{w}_a(x) \hat{w}_b(y)
{E}^a(x) {E}^b(y) f_\epsilon(x-y)
\end{equation}
where $\lim_{\epsilon\rightarrow 0} f_\epsilon(x-y) = \delta(x-y)$, so
it is a usual point-splitting regulator, and we have eliminated the
tildes to simplify notation, and as we stated above, treat the
electric fields as classical quantities. The operators $\hat{w}_a$
only act at intersections of the weave, so the integrals are replaced
by discrete sums when evaluating the action of the Hamiltonian on a
weave state,
\begin{equation}
<\Delta|\hat{H}^E_{\rm Maxwell}|\Delta>   = {1 \over 2} 
\sum_{v_i,v_j } 
<\Delta|\hat{w}_a(v_i) \hat{w}_b(v_j) |\Delta>
{E}^a(v_i) {E}^b(v_j) 
\end{equation}
where $v_i$ and $v_j$ are vertices of the weave and the summation
includes all vertices within the domain of characteristic length
$\Delta$.  We now expand the electric field around the central point of the
$\Delta$ domain,  which we call $P$, and get,
\begin{equation}\label{expansion}
E^a(v_i) \sim E^a(P) + (v_i-P)_c \partial^c E^a(P)+\cdots,
\end{equation}
and given the assumptions we made about the long wavelength nature of
the electric fields involved, we will not need to consider higher
order terms in the expansion at the moment. Notice that $(v_i-P)_c$ is
a vector of magnitude approximately equal to $\Delta$, whereas the
partial derivative of the field is of order $1/\lambda$, that is, we
are considering an expansion in $\Delta/\lambda$. We now insert this
expansion in the Hamiltonian and evaluate the resulting terms in the
weave approximation. One gets two types of terms, one is given by the
product of two electric fields evaluated at $P$ times the sum over the
vertices of the metric operator. Due to the definition of the weave state, 
the sum just yields the classical metric and we recover the usual 
Maxwell Hamiltonian in flat space. 

We now consider the next terms in the expansion $\Delta/\lambda$. They
have the form,
\begin{equation}
{1 \over 2} 
\sum_{v_i,v_i} 
<\Delta|\hat{w}_a(v_i) \hat{w}_b(v_j) |\Delta> (
(v_i-P)_c 
\partial_c ({E}^a(P)) {E}^b(P) +
(v_j-P)_c
{E}^a(P) \partial_c ({E}^b(P)),
\end{equation}
 When performing the sum over all vertices in the cell we discussed
above, we end up evaluating the quantity $<~\Delta|\hat{w}_a(v_i)
\hat{w}_b(v_j) |\Delta> (v_i-P)_c$.  This quantity averages out to
zero in a first approximation, since one is summing over an isotropic
set of vertices. The value of the quantity is therefore proportional
to $\ell_P/\Delta$, the larger we make the box of characteristic
length $\Delta$ the more isotropic the distribution of points is. We
consider the leading contribution to this term, which should be a
rotational invariant tensor of three indices, i.e., it is given by
$\chi \epsilon_{abc} \ell_P/\Delta$ with $\chi$ a proportionality
constant of order one (that can be positive or negative).

We have therefore found a correction to the Maxwell Hamiltonian
arising from the discrete nature of the weave construction. It should
be noticed that the additional term we found is rotationally
invariant, i.e., it respects the original spirit of the weave
construction. It is, however, parity violating. If one were to assume
that the weaves are parity-invariant, the term would vanish. The term
would also vanish ---on average--- if one assumes that the different
regions of size $\Delta$ have ``random orientations'' in their parity
violation. The fact that we live in a non-parity invariant universe
suggests that parity invariant weaves might not necessarily be the
most natural ones to consider in constructing a semiclassical state of
cosmological interest.

Assuming a non-parity invariant weave, the resulting equations of
motion from the above Hamiltonian can be viewed as corrections to the
Maxwell equations,
\begin{eqnarray}
\partial_t \vec{E} &=& -\nabla \times \vec{B} + 2 \chi {\ell_P}
\Delta^2 \vec{B}\\
\partial_t \vec{B} &=& \nabla \times \vec{E} - 2 \chi {\ell_P}
\Delta^2 \vec{E}.
\end{eqnarray}
As we see the equations gain a correction proportional to the
Laplacian $\Delta^2$ of the fields, the correction is symmetrical in
both fields, but is not Lorentz covariant. This already suggests that
there will be modifications to the usual dispersion relation for light
propagation. The lack of covariance is not surprising, since the weave
selects a preferred foliation of spacetime. This again is what is
standardly accepted in cosmological applications as we will consider,
there is a preferred set of comoving observers, and for such observers
we will compute the effect to be observed.

If one now combines the above equations to study wave propagation, 
we get,
\begin{equation}
\partial_t^2 \vec{E} - \Delta^2 \vec{E} -4\chi {\ell_P}
\Delta^2 (\nabla \times \vec{E})
\end{equation}
and similarly for $\vec{B}$. We now seek solutions with a given
helicity, 
\begin{equation}
\vec{E}_\pm = {\rm Re}\left((\hat{e}_1 \pm i\hat{e}_2)
e^{i(\Omega_\pm t - \vec{k}\cdot\vec{x})}\right).
\end{equation}
Substituting in the above equations, we get 
\begin{equation}
\Omega_\pm = \sqrt{k^2 \mp 4 \chi\ell_P k^3} \sim |k| (1 \mp 2 \chi
\ell_P |k|).
\end{equation}

We therefore see the emergence of a birefringence effect, associated
with quantum gravity corrections. The group velocity has two branches,
and the effect is of the order of a shift of one Planck length per
wavelength.

This effect is distinct from other effects that have been discussed in
the past. If we compare with the proposals considered by
Amelino-Camelia et al \cite{Amelino}, in their case they find
only a change in the dispersion relation, whereas here we in addition
see a helicity-dependent effect. Our effect is also absent for scalar
fields, whereas other quantum gravity corrections are all-encompassing
(they can be viewed as corrections to quantum mechanics
itself). Birefringence was also considered in the context of
modifications of electromagnetism and also in nonsymmetric gravity
\cite{haugan}. In those cases the effect was not frequency
dependent. This is because the kind of corrective terms we are
introducing in Maxwell theory, although linear in the fields, are
higher order in the derivatives. This is in line with the observations
of ref. \cite{Amelino}, that quantum gravity effects will increase
with the frequency, the opposite being expected for other more
standard sources of cosmological dispersion or birefringence.

To quantify the magnitude of these effects, if one considers a
gamma-ray-burster at cosmological distances (about $10^{10}$ light
years) and frequencies of the order of $200keV$ (like the channels of
the BATSE detector), this implies a delay between the two group
velocities of both polarizations that compose a plane wave of
$10^{-5}s$. The observed width of the bursts appears to be of the
order of $0.1s$, with features like a rising edge as small as $1 ms$.
We therefore see that with such observations one is two orders of
magnitude away of observing these effects. This is fairly impressive
given that this is an effect due to quantum gravity. The intention of
this note is not to present a detailed calculation of the magnitude of
the effects, however, one could envision a more subtle program to seek
for the effect, given its distinctive signature, and its specific
dependence on frequency, using data from more than one channel and
more sophisticated pattern matching techniques. 

How did a birefringence appear? In the construction of the weave, we
have assumed that rotational invariance is locally preserved. However,
we have not assumed that parity invariance is preserved, and in the
model considered it is violated. That is, one can envisage a
fundamental, Planck-level violation of parity in the weave approach,
without detriment to the ability of the weave to approximate a given
metric. Which weave to choose (parity preserving or violating) is a
reasonable issue to settle experimentally. The measurements of spectra
of gamma-ray bursts might provide a mechanism for this. It is
intriguing to see if other symmetries might be violated and which
observational consequences it might have. It is in this sense that
this paper can be viewed as further conditions that must be met by the
semi-classical states of the theory.

In general, without further input from the dynamics of the theory, one
would expect that a weave structure would lead to the loss of
Poincar\'e invariance of the Maxwell equations. In the example
considered we see that this invariance is broken simultaneously with
parity invariance. It is interesting to notice that if the weave is
parity preserving Poincar\'e invariance is preserved as well.

Another viewpoint could be that if at some point a complete dynamical
theory is established that determines the evolution of the weaves, one
could presumable construct quantum toy cosmological models. In such
situation the final weave describing our current universe would be
prescribed and one could determine if the theory predicts the presence
of birefringence or not along the lines discussed in this paper.

Finally, at a more formal level, the appearance of corrections to the
propagation of light might allow to study effects concerning
information loss in black hole systems.  These considerations are
currently under study.

We wish to thank Abhay Ashtekar and Mike Reisenberger for various
insightful comments. This work was supported in part by grants
NSF-INT-9406269, NSF-PHY-9423950, research funds of the Pennsylvania
State University, the Eberly Family research fund at PSU and PSU's
Office for Minority Faculty development. JP acknowledges support of
the Alfred P. Sloan foundation through a fellowship. We acknowledge
support of  PEDECIBA (Uruguay).

\end{document}